\documentclass[preprint,review,12pt]{elsarticle}

\usepackage{amsmath}
\usepackage{amssymb}
\usepackage{amsfonts}
\usepackage{graphicx}
\usepackage{dcolumn}
\usepackage{bm}
\usepackage{hyperref}
\hypersetup{colorlinks=false}
\biboptions{numbers,square}

\begin{document}

\begin{frontmatter}

\title{Casimir Effect in the Ho\v{r}ava-Lifshitz Gravity with a Cosmological Constant}

\author{C. R. Muniz}

\address{Grupo de F\'isica Te\'orica (GFT), Universidade Estadual do Cear\'a, UECE-FECLI, Iguatu-CE, Brazil.}

\ead{celiomuniz@yahoo.com}

\author{V. B. Bezerra}

\address{Departamento de F\'{i}sica, Universidade Federal da Para\'{i}ba, Caixa Postal 5008, CEP 58051-970, Jo\~{a}o Pessoa-PB, Brazil}

\ead{valdir@ufpb.br}

\author{M. S. Cunha}

\address{Grupo de F\'isica Te\'orica (GFT), Universidade Estadual do Cear\'a-UECE, CEP 60714-903, Fortaleza-CE, Brazil}

\ead{marcony.cunha@uece.br}

\begin{abstract}
We calculate the Casimir energy of a massless scalar field confined between two nearby parallel plates formed by ideal uncharged conductors, placed tangentially to the surface of a sphere with mass $M$ and radius $R$. To this end, we take into account a static and spherically symmetric solution of Ho\v{r}ava-Lifshitz (HL) gravity, with a cosmological constant term, in lower orders of approximation, considering both weak-field and infrared limits. We show that the Casimir energy, just in the second order weak-field approximation, is modified due to the parameter of the HL gravity as well as to the cosmological constant.
\end{abstract}


\end{frontmatter}


\section{Introduction}

The theoretical formulation known as Ho\v{r}ava-Lifshitz gravity is a recent attempt
to construct a renormalizable theory of gravity \cite{horava1,horava2,horava3} which reduces
to general relativity in the appropriate limit. This formulation exhibits a broken Lorentz symmetry in ultraviolet (UV) regime, which manifests itself in a strong anisotropic scaling of space and time, namely, $t\rightarrow b^zt$ and $x^{i}\rightarrow bx^{i}$, where $b$ is an arbitrary scale factor and $z$ is a dynamical critical exponent that explicitly breaks the Lorentz symmetry. The validity of the general relativity must be restored at infrared (IR) scale, or at least Lorentz violations in this scale are requested to stay below current experimental constraints. In $(3+1)$-dimensional spacetime, the renormalizability of the theory exists for $z=3$, and, in general, it occurs for $z=d$ in a $(d+1)$-dimensional spacetime.

Ho\v{r}ava-Lifshitz gravity has smaller strength at high energy scale than in general relativity, due to the presence of the $z$ order spatial derivatives in the action, and is ghost-free by demanding only first order temporal derivatives. Such features generate a bouncing cosmology and an accelerating universe, {\it i.e}, dark energy \cite{mukohyama}. Furthermore, the cosmological constant present in Ho\v{r}ava theory is very large and negative, and can solve the problem of huge discrepancy between the observed value of the cosmological constant and the predictions from quantum field theory, which indicates a large and positive value for the global vacuum energy of matter \cite{corrado}. This new theory of gravity, inspired by dynamical critical systems occurring in condensed matter physics is compatible with other traditional approaches to construct a renormalizable quantum theory of gravity, such as causal dynamical triangulations (CDT) \cite{ambjorn} and renormalization group (RG) approaches based on asymptotic safety \cite{niedermaier}.

While the cosmological constant is a manifestation of the vacuum in global scales, the Casimir effect arises locally, being originally a physical phenomenon which consists in an attractive force occurring between two parallel idealized uncharged conductors placed in vacuum \cite{casimir1}. This is a purely quantum effect due to the modifications of the zero-point oscillations of the electromagnetic field and results from the presence of material boundaries in comparison with Minkowski space. This phenomenon strongly depends on the geometry of the boundaries \cite{milton, bordag, milonni} among others aspects, such as the geometry and topology of the manifold.

Investigations on Casimir effect in a gravitational field, in weak-field approximation and in the context of the general relativity, were carried on in a series of papers \cite{milton1,milton2,milton3,fulling}, which presented interesting discussions about the gravitational Casimir energy and the equivalence principle. As was noticed in the 1970's \cite{DeWitt,Ford}, the Casimir effect arises also in empty spaces with non-trivial topology. In such spaces, there are no material boundaries, but instead of these, there are some identification conditions imposed by the topology on quantum fields, which play the role as the ones imposed by material boundaries, modifying the spectrum of the vacuum oscillations as compared to the Minkowski spacetime. This allows that the Casimir effect becomes a global phenomenon, even a cosmological one (\cite{Herondy}, and references therein).

In a recent paper \cite{celio}, the Casimir effect was examined in the background gravitational field corresponding to a static and spherically symmetric black hole solution \cite{kehagias} of Ho\v{r}ava-Lifshitz gravity. This solution represents a particular case of a general solution that corresponds to a black hole with a cosmological constant \cite{Park}, which reduces, in the IR limit, to the standard AdS black hole solution obtained in \cite{Lu} as well as to the Schwarzschild solution when $\omega\rightarrow\infty$, where $\omega$ is a free parameter that regulates the IR behavior of the Horava-Lifshitz gravity.

In a general sense, it is important to take into account the possible Ho\v{r}ava-Lifshitz gravity modifications on the Casimir effect in the presence of gravitational fields, because this effect gives rise to some nonzero stress-energy tensor of a quantum field in the vacuum state which depends on the geometrical parameters of the manifold.

In this paper we complete the calculations done in \cite{celio}, for $T=0$, in the sense that we are considering a solution which represents a generalization of that one obtained in \cite{kehagias}, by including an arbitrary cosmological constant \cite{Park}. Then, we calculate the Casimir energy up to second order weak-field approximation following the procedure used in \cite{sorge}.

The paper is organized as follows. In section 2, we present the gravitational
background to be used, in the context of Ho\v{r}ava-Lifshitz gravity. In section 3, we calculate the
regularized vacuum energy (Casimir energy) of the massless scalar field in this
background, in first order of both weak field and IR approximations. In section 4,
we do similar calculations in the next order of approximation, finding the modifications in the
the Casimir energy induced by the gravitational field under consideration. Finally, in section 5, we discuss the results.

\section{THE STATIC AND SPHERICALLY SYMMETRIC SOLUTION IN  HO\v{R}AVA-LIFSHITZ GRAVITY WITH A COSMOLOGICAL CONSTANT}

To begin with, let us write the static and spherically symmetric general solution of Ho\v{r}ava-Lifshitz gravity with a cosmological constant, whose line element is given by \cite{Park}
\begin{eqnarray}\label{01}
ds^2=f(r)dt^2-\frac{1}{f(r)}dr^2-dr^2-r^2(d\theta^2+\sin^2{\theta}d\phi^2),
\end{eqnarray}
where
\begin{equation}\label{02}
f(r)=1+\left(\omega-\frac{2}{3}\Lambda\right)r^2-\sqrt{r\left[\omega\left(\omega-\frac{4}{3}\Lambda\right)r^3+\beta\right]},
\end{equation}
with $\beta$ and $\Lambda$ being an integration constant and the cosmological constant, respectively. Choosing $\beta=4M\omega$, where $\omega$ is a free parameter that regulates the UV behavior of the theory, and $\Lambda=0$, we recover the solution found by Kehagias and Sfetsos \cite{kehagias}, which reduces to the Schwarzschild solution when one takes the IR limit $\omega\rightarrow\infty$,  .

Considering $\beta=4M\omega$, we can rewrite Eq. (\ref{02}), in a more appropriate form, given by
\begin{equation}\label{03}
f(r)=1+\omega r^2\left(1-\frac{2\Lambda}{3\omega}\right)-\omega
r^2\sqrt{1-\frac{4\Lambda}{3\omega}+\frac{4M}{\omega r^3}}.
\end{equation}

Expanding the square root up to first order in both $\Lambda/\omega$ and
$\Lambda/\omega r^3$, we reobtain the Schwarzschild metric coefficients
$f^{1,1}(r)\approx 1-2M/r$, where the superscripts mean the orders of approximation.
In this background the Casimir energy associated to the massless scalar field was already found by Sorge \cite{sorge}.

Now, let us consider an expansion of Eq. (\ref{03}) up to a subsequent order, namely, $\mathcal{O}(\Lambda/\omega)^2$,
keeping terms of $\mathcal{O}(\Lambda/\omega r^3)$ and $\mathcal{O}(M/r)$. Doing this we find that the metric
coefficients are given by
\begin{equation}\label{04}
f^{2,1}(r)=1-\frac{2\widetilde{M}}{r}+\frac{\widetilde{\Lambda}r^2}{3},
\end{equation}
where the superscripts $1$ and $2$ mean the first and second orders of approximation. Note that Eq. (\ref{04}) corresponds
to an effective Schwarzschild-AdS metric, with
$\widetilde{M}=M\left(1+\frac{2\Lambda}{3\omega}\right)$ and
$\widetilde{\Lambda}=\frac{2\Lambda^2}{3\omega}$.

Due to the characteristics of the problem under consideration, it is more convenient to use isotropic coordinates, which allows to introduce the geometry of the parallel plates in a simple way.
\begin{figure}[!h]
\includegraphics[scale=0.9]{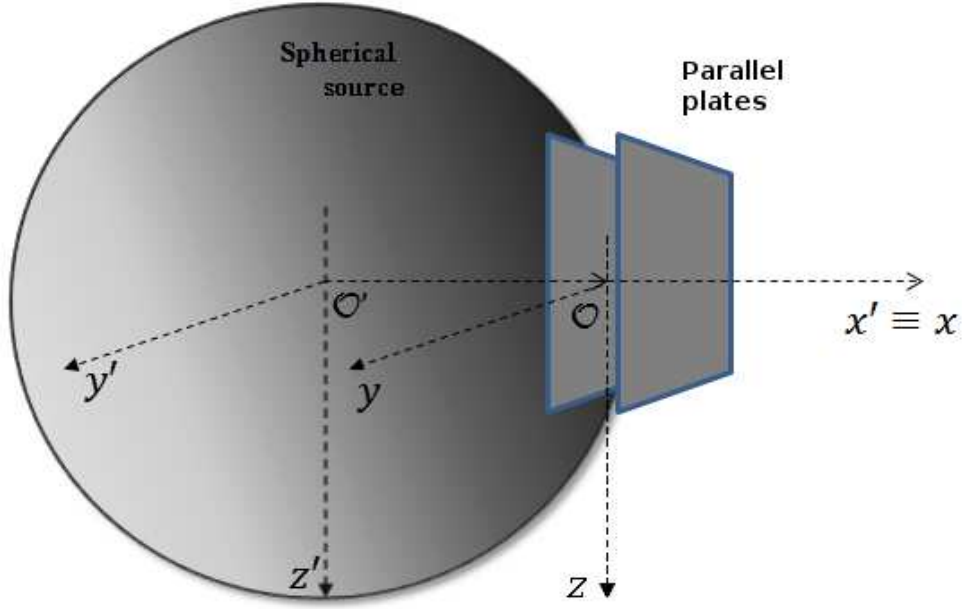}
\caption{Casimir energy, in units of Planck mass, $M_P$, for the massive scalar field around a spherical source with $M=10 M_P$, as a function of the mass of the field, $m$, in units of $M_P$.}
\end{figure} 
 
In Fig. 1, out of scale, we depict the disposition of the plates with respect to the spherical source and the coordinates used. The Cartesian coordinates $(x, y, z)$ have origin on the first plate which is tangent to the source. They are related to the coordinates 
$x'$, $y'$, $z'$, whose origin coincides with the center of the source by: $x=x'-R$, $y=y'$, $z=z'$. It is worth calling attention to the 
fact that $r^2=x'^2+y'^2+z'^2\simeq x'^2$, and then in the first order of approximation, we can consider $r=x'\simeq R$, with $R$ being the radius of the gravitational source. Thus, taking into account the metric coefficients given by Eq. (\ref{04}), we obtain the following line element 
\begin{equation}\label{05}
ds^2=\left(1+2\widetilde{\Phi}_0+\frac{\widetilde{\Lambda}R^2}{3}\right)dt^2-\left(1-2\widetilde{\Phi}_0+\frac{\widetilde{\Lambda}R^2}{6}\right)(dx^2+dy^2+dz^2),
\end{equation}
The potential ${\Phi}_0$ is given by
\begin{eqnarray}\label{06}
\widetilde{\Phi}_0=-\frac{\widetilde{M}}{R}
\end{eqnarray}

Then, we will take into account the second order term in $(M/\omega r^3)$ and first order term in $(\Lambda/\omega)$, in the Eq. (\ref{03}), such that we will get
\begin{equation}\label{07}
f^{1,2}(r)=1-\frac{2\widetilde{M}}{r}+\frac{2\widetilde{M}^2}{\widetilde{\omega}r^4},
\end{equation}
where $\widetilde{\omega}=\omega\left(1-\frac{2\Lambda}{3\omega}\right)$. Thus, the resulting metric can be written as
\begin{equation}\label{08}
ds^2=\left(1-\frac{2\widetilde{M}}{r}+\frac{2\widetilde{M}^2}{\widetilde{\omega}r^4}\right)dt^2-\left(1+\frac{2\widetilde{M}}{r}-\frac{\widetilde{M}^2}{2\widetilde{\omega}r^4}\right)(dx^2+dy^2+dz^2),
\end{equation}
where the cosmological constant term is included in the definition of $\widetilde{M}$.

\section{CASIMIR EFFECT IN THE FIRST ORDER WEAK-FIELD APPROXIMATION}

In this section we will calculate the regularized vacuum energy of a massless scalar field in the first order approximation. It is worth noticing that the results which will be presented in this section can be obtained more directly by means of an adequate transformation of coefficients of the metric given in Eq. (\ref{05}). However, in order to show the results, step by step, corresponding to the first order of approximation, we will present the complete calculation to fix the methodology, which will must be used in the following order of approximation.
Thus, we will consider the Klein-Gordon equation for the massless scalar field in the background given by Eq. (\ref{05}), which can be written as
\begin{equation}\label{09}
\frac{1}{\sqrt{-g}}\partial_{\mu}(\sqrt{-g}g^{\mu\nu}\partial_{\nu})\Psi=\left(1-4\widetilde{\Phi}_0-\frac{\widetilde{\Lambda}
R^2}{6}\right)\partial_t^2\Psi-\nabla^{2}\Psi=0,
\end{equation}
where $\nabla^{2}\equiv\partial_x^2+\partial_y^2+\partial_z^2$.

Let us assume that the solutions of Eq. (\ref{09}) are given by
\begin{equation}\label{10}
\Psi_{n}(x,y,z,t)=N_{n}\exp{[i( k_yy+k_zz-\Omega_{n} t)]}\sin{(n\pi x/L)}
\end{equation}
with the Dirichlet boundary conditions being satisfied on the plates. Substituting
Eq. (\ref{10}) into (\ref{09}), we find that the eigenfrequencies $\Omega_{n}^{2,1}$
are
\begin{equation}\label{11}
\Omega_{n}^{2,1}=\left(1+2\widetilde{\Phi}_0+\frac{\widetilde{\Lambda}
R^2}{12}\right)\left(k_y^2+k_z^2+\frac{n^2\pi^2}{L^2}\right)^{1/2}
\end{equation}

We proceed by calculating the normalization constant $N_n$, in order to find the
vacuum expectation value of the $00$ component of the energy-momentum tensor. To do
this, we will consider the norm  of the scalar functions $\Psi_n$ (which obey the
usual orthonormality conditions), defined on the spacelike Cauchy hypersurface,
$\Sigma$, as \cite{birrel},
\begin{equation}\label{12}
\|\Psi_n\|=-i\int_{\Sigma}\sqrt{-g_{\Sigma}}[\partial_t(\Psi_n)\Psi_n^{*}-\partial_t(\Psi_n^{*})\Psi_n]n^{t}d\Sigma
\end{equation}
where $g_{\Sigma}$ is the determinant of the metric induced on the hypersurface,
$g_{ik}$, with $i,k = 1,2,3$ and $d\Sigma=dxdydz$ is its volume element. Defining
the timelike future-directed unitary vector ${\bf n^t}$ as
\begin{equation}\label{13}
{\bf n^t}\equiv(n^t,0,0,0)=\left(1-\widetilde{\Phi}_0-\frac{\widetilde{\Lambda}
R^2}{6},0,0,0\right),
\end{equation}
we obtain using Eqs. (\ref{13}) and (\ref{12}) that the square of normalization
constant is given by
\begin{equation}\label{14}
N^2_{n}=\frac{1}{(2\pi)^2\left(1-4\widetilde{\Phi}_0+\frac{\widetilde{\Lambda}
R^2}{12}\right)L\Omega_{n}}.
\end{equation}
The vacuum expectation value of the energy density $T_{tt}$ of the scalar field is
\begin{equation}\label{15}
\epsilon^{2,1}_{vac}=<0|n^tn^tT^{2,1}_{tt}(\Psi_{n})|0>=\int
d^2k_{||}\sum_nn^tn^tT^{2,1}_{tt}(\Psi_n)
\end{equation}
where $d^2k_{||}=dk_ydk_z$ and
\begin{eqnarray}\label{16}
T^{2,1}_{tt}(\Psi_n)&=&\frac{1}{2}\left(\partial_t\Psi_n\partial_t\Psi^{*}_n-g_{tt}g^{ik}\partial_i\Psi_n\partial_k\Psi^{*}_n\right)\nonumber\\
&=&N^2_n\left(1+4\widetilde{\Phi}_0+\frac{\widetilde{\Lambda}
R^2}{6}\right)\left[k_{||}^2\sin^2{\left(\frac{n\pi
x}{L}\right)}+\frac{n^2\pi^2}{2L^2}\right].
\end{eqnarray}
In order to obtain the vacuum energy density $\overline{\epsilon}^{2,1}_{vac}$, we take the average value of
$\epsilon^{2,1}_{vac}$ in the spatial region (cavity) delimited by the plates, which is defined by
\begin{equation}\label{17}
\overline{\epsilon}_{vac}=\frac{1}{V_p}\int_{\Sigma}\sqrt{-g_{\Sigma}}\epsilon^{2,1}_{vac}d\Sigma
\end{equation}
where $V_p=\int_{\Sigma}\sqrt{-g_{\Sigma}}d\Sigma$ is the proper volume of the cavity.

Putting (\ref{16}) into (\ref{15}) and substituting the obtained result into (\ref{17}), we get
\begin{eqnarray}\label{18}
\overline{\epsilon}^{2,1}_{vac}=\left(1+4\widetilde{\Phi}_0-\frac{\widetilde{\Lambda}
R^2}{3}\right)\int\frac{d^2k_{||}}{2(2\pi)^2L}\sum_n\left[k^2_{||}+\left(\frac{n\pi}{L}\right)^2\right].\nonumber\\
\end{eqnarray}
The integral in the rhs of (\ref{18}) is just the vacuum energy density of the
massless scalar field in Minkowski space, which is infinity. The renormalized Casimir energy
density $\overline{\epsilon}^{ren}_{vac}$ is found by using an adequate renormalization technique, as the Abel-Plana subtraction formula \cite{bordag}, which we will consider in here. Applying this renormalization process in equation (\ref{18}), we obtain the following result
\begin{equation}\label{19}
\overline{\epsilon}^{ren,(2,1)}_{vac}=-\left(1+4\widetilde{\Phi}_0-\frac{\widetilde{\Lambda}
R^2}{3}\right)\frac{\pi^2}{1440L^4}.
\end{equation}

If we take into account the proper length
$L_p=\left(1-\widetilde{\Phi}_0+\frac{\widetilde{\Lambda} R^2}{12}\right)L$, we find that the Casimir energy given by Eq. (\ref{19}) becomes
\begin{eqnarray}\label{20}
\overline{\epsilon}^{ren,(2,1)}_{vac}=-\frac{\pi^2}{1440L_p^4},
\end{eqnarray}
which is the result obtained in the Minkowski spacetime. Notice that independently of both the values of $\omega$ and $\Lambda$, we recover the result obtained by Sorge \cite{sorge}, which means that the Casimir energy is the same as the one obtained in flat spacetime, a fact that is consistent the constant gravitational potential taken into account.
As we pointed out, this is an expected result which can be obtained just by redefining the coordinates in the background spacetime under consideration.

In the general case, for arbitrary values of $\omega$ and $\Lambda$, the proper distance $L_p$ is affected by these parameters in such way that the Casimir energy would be different. In the case examined in this section, for the approximation considered, gravitation act as a static and uniform potential, and for this reason we obtained a result similar as the one in Minkowski spacetime \cite{sorge}, as it should be expected.

\section{The Casimir effect in the second order weak-field approximation}

Now, let us consider similar calculations to those ones made in the previous section to find the regularized vacuum energy of the massless scalar field, but now in the second order weak-field approximation. To proceed with the calculations, let us consider the changing $x'\simeq r=R+x$, with $x\ll R$, and substitute it into Eq. (\ref{08}), remembering that we have assumed the origin of the coordinate system on one of the plates. Note that only linear terms in $(x/R)$ will be taken into account. Therefore, we can write Eq. (\ref{08}) in the following form
\begin{equation}\label{21}
ds^2\simeq(1+2b\gamma x)dt^2-(1-2\gamma x)(dx^2+dy^2+dz^2),
\end{equation}
where $b=1-\frac{3\widetilde{M}}{\widetilde{\omega} R^3}$ and
$\gamma=\left(\frac{\widetilde{M}}{R^2}-\frac{\widetilde{M}^2}{\widetilde{\omega}
R^5}\right)$.

The equation for a massless scalar field in the spacetime with the metric given by (\ref{21}) is
\begin{equation}\label{22}
[1-2(b+1)\gamma x]\partial_t^2\Psi-\nabla^{2}\Psi-(b-1)\gamma\partial_x\Psi=0.
\end{equation}

Let us consider that the solutions of the equation (\ref{22}) can be written in the form
\begin{equation}\label{23}
\Psi_{n}(x,y,z,t)=N_n\exp{[i(k_yy+k_zz-\Omega_n t)]}\chi (x),
\end{equation}
where $\chi(x)$ is a function to be determined. Substituting Eq. (\ref{23}) into Eq.
(\ref{22}), we get
\begin{equation}\label{24}
\frac{d^2\chi}{dx^2}-(b-1)\gamma\frac{d\chi}{dx}-4\widetilde{\gamma}\omega^2x\chi+(\Omega_n^2-k_{||}^2)\chi=0
\end{equation}
where
\begin{equation}\label{25}
\widetilde{\gamma}=\frac{\gamma(b+1)}{2}
\end{equation}
The general solution $\chi(x)=C_1\chi^{(1)}+C_2\chi^{(2)}$ is given in terms of the Airy functions
$\text{Ai(x)}$ and $\text{Bi(x)}$  in the form
\begin{eqnarray}\label{26}
\chi^{(1)}(x)= \exp{\left[\frac{-\gamma(1-b)x}{2}\right]}\text{Ai}\left[\frac{k_{||}^2-\Omega_n^2+4 \widetilde{\gamma} \Omega_n^2
x-\gamma^2(b-1)^2}{\left(4\widetilde{\gamma} \Omega_n^2\right)^{2/3}}\right].\\
\chi^{(2)}(x)= \exp{\left[\frac{-\gamma(1-b)x}{2}\right]}\text{Bi}\left[\frac{k_{||}^2-\Omega_n^2+4 \widetilde{\gamma} \Omega_n^2
x-\gamma^2(b-1)^2}{\left(4\widetilde{\gamma} \Omega_n^2\right)^{2/3}}\right].
\end{eqnarray}
The term $\gamma^2(b-1)^2$ in the numerator of the Airy function argument is negligible due to the approximations used. On the other hand, the remaining term
\begin{equation}
z(x)=-\frac{(\Omega_n^2-k^2_{||}-4\widetilde{\gamma}\Omega_n^2x)}{(4\widetilde{\gamma}\Omega^2_n)^{2/3}},
\end{equation}
is very large since that $\widetilde{\gamma}\ll1$. Negative values of $z$ are not physically compatible with the Dirichlet boundary conditions because of the asymptotic behaviour of the Airy functions in that domain. Thus we can use the asymptotic expansion for large values of the Airy functions Ai($-z$) and Bi($-z$), for $z>0$ , which can be combined in a such way that \cite{abramowitz}
\begin{equation}\label{27}
\chi(x)\sim C\exp{\left[\frac{-\gamma(1-b)x}{2}\right]}
\frac{\sin{(\frac{2}{3}z^{3/2}+\phi})}{\sqrt{\pi}z^{1/4}},
\end{equation}
where $\phi$ is a constant phase to be determined by the boundary conditions on the plates.
Considering the Dirichlet boundary conditions $\chi(0)=\chi(L)=0$, then we have
\begin{equation}
z^{3/2}(0)-z^{3/2}(L)\thickapprox\frac{3n\pi}{2},
\end{equation}
with $n=0,1,2,3...$. Expanding the left hand side of the above equation up to second order in $4\widetilde{\gamma}\Omega_n^2L/(\Omega^2_n-k^2_{||})$, we can find the following eigenfrequencies
 \begin{equation}\label{28}
 \Omega_{n,2}^2=(1+2\widetilde{\gamma}
L)\left(k_{||}^2+\frac{n^2\pi^2}{L^2}\right)=[1+\gamma(b+1)L]\left(k_{||}^2+\frac{n^2\pi^2}{L^2}\right),
 \end{equation}
where the subscript 2 means that we are considering the weak-field second order of approximation.

In order to calculate the quantum vacuum energy for this order of approximation,
let us assume that the solution of Eq. (\ref{24}) can also be written as a
correction of the harmonic modes obtained in the case when the post-Newtonian
parameter is expanded up to $\mathcal{O}(M/R)$.

Let us represent the solution corresponding to the second order as $\Psi_{n,2}$, where the subscript $2$ means the order of
approximation. Thus, we can write
\begin{equation}\label{29}
\Psi_{n,2}=\Psi_{n,1}+\delta\Psi_{n,1},
\end{equation}
where $\Psi_{n,1}$ is given by Eq. (\ref{10}). Therefore, the expression for
the quantum vacuum energy can be splited as
\begin{equation}\label{30}
\overline{\epsilon}_{vac,2}=\overline{\epsilon}_{vac}+\delta\overline{\epsilon}_{vac}
\end{equation}
where
\begin{equation}\label{31}
\overline{\epsilon}_{vac}=\frac{1}{V_p}\int
d^2k_{||}\sum_n\int_{\Sigma}\sqrt{-g_{\Sigma}}n^tn^t T_{tt}(\Psi_{n,1})d\Sigma.
\end{equation}
Employing the same procedure already used to calculate the quantum
vacuum energy and taking into account that, in the present situation,
$n^t\simeq1-b\gamma x$, we get
\begin{equation}\label{32}
\overline{\epsilon}_{vac}=\left[1+(b-1)\gamma L\right]\int \frac{1}{2(2\pi)^2L}
d^2k_{||}
\sum_n \Omega_{0,n},
\end{equation}
where $\Omega_{0,n}=\left(k^2_{||}+\frac{n^2\pi^2}{L^2}\right)$. The renormalization
of the integral which appears in Eq. (\ref{29}) results in the flat Casimir energy
density given by
 \begin{equation}\label{33}
\overline{\epsilon}^{ren}_{vac}=-\left[1+(b-1)\gamma L\right]\frac{\pi^2}{1440 L^4}.
 \end{equation}
Now, our task is to calculate $\delta\overline{\epsilon}_{vac}$ which can be written as
 \begin{equation} \label{34}
 \delta\overline{\epsilon}_{vac}=\frac{1}{2(2\pi)^2L}\int
d^2k_{||}\sum_n(\Omega_{2,n}-\Omega_{0,n})=\frac{(b+1)\gamma
L}{2}\frac{1}{2(2\pi)^2L}\int d^2k_{||}\sum_n\Omega_{n,0}.
 \end{equation}
 Finally, taking into account the proper length obtained from the relation $L\simeq[1+(1/2)\gamma L_p]L_p$,
the second order correction to the renormalized Casimir energy is

 \begin{equation}\label{35}
 \overline{\epsilon}^{ren}_{vac,2}=\overline{\epsilon}^{ren}_{vac}+\delta\overline{\epsilon}^{ren}_{vac}=
-\left[1-\frac{L_p\widetilde{M}}{R^2}\left(1
-
\frac{7\widetilde{M}}{2\widetilde{\omega}R^3}\right)\right]\frac{\pi^2}{1440L_p^4}.
 \end{equation}

Notice that this result corresponds to the Minkowski spacetime one rescaled by a factor which depends on the Ho\v{r}ava-Lifshitz parameter and on the cosmological constant. It is worth calling attention to the fact that the background geometry is not homogeneous, but the Casimir energy is a global quantity, only depending on the geometrical parameter which corresponds to the separation between the plates, $L$, and does not depend explicitly on the locations of them. In other words, the Ho\v{r}ava-Lifshitz gravity scenario as well as the presence of the cosmological constant induces an effective modification in the Casimir energy in the weak-field second order of approximation. It is worth noting that when $\Lambda=0$ and $\omega\rightarrow\infty$, we get the same result obtained by Sorge \cite{sorge}.

\section{Concluding remarks}

We have calculated the Casimir energy of a massless scalar field in two nearby,
parallel, uncharged conductor plates placed tangentially to the surface of a sphere
with mass $M$ and radius $R$, at zero temperature. The analysis has taken into
account the effects due to a static, spherically symmetric gravitational
background with a cosmological constant, in the context of the
Ho\v{r}ava-Lifshitz gravity. For this analysis, we considered a solution described by
Park \cite{Park} of the vacuum field equations in the Ho\v{r}ava-Lifshitz gravity, which taken into account
an arbitrary cosmological constant.

The obtained results show us that when $\omega\rightarrow\infty$, which corresponds to the IR limit ($\omega\gg1$), and in the first order weak-field approximation, independently of the value of $\Lambda$, we recover the result obtained by Sorge \cite{sorge}, meaning that the Casimir energy is the same as the one obtained in flat spacetime. This is consistent with the fact that in this approximation, we considered a static and uniform gravitational potential. In this same approximation, for arbitrary values of $\omega$ and $\Lambda$, the proper distance $L_p$ is affected by these parameters, which means that the Casimir energy is also influenced by them, namely, $\omega$ and by $\Lambda$. Otherwise, as we have a situation which corresponds to a static and uniform gravitational potential, the result is formally similar to one obtained in Minkowski spacetime \cite{sorge}, as it should be expected.

In the following order of approximation, the difference between the results obtained in the Ho\v{r}ava-Lifshitz gravity and in general relativity is far from the formal similarity. This can be seen clearly by analysis of the result expressed by Eq. (\ref{35}). From this equation, we can conclude that there is an attenuation of the Casimir force for a given relationship between the cosmological constant and HL gravity parameter, namely, if $\Lambda<\frac{3\omega^2}{4}$ (remind that we are working in the IR limit, for which $\omega\gg1$). Otherwise, an amplification occurs. However, in both the situations, the attractive character of the Casimir force is preserved.

\section*{ACKNOWLEDGMENTS}

C.R. Muniz would like to thank Universidade Federal da Para\'iba (UFPB) for the kind
welcome. V.B.Bezerra would like to thank
CNPq for partial financial support.

\end{document}